\documentclass[12pt,preprint]{aastex}
\begin{document}
\title{GALEX and Optical Light Curves of EF Eridanus During a Low State: the Puzzling Source of UV Light}

\author{Paula Szkody\altaffilmark{1},
Thomas E. Harrison\altaffilmark{2},
Richard M. Plotkin\altaffilmark{1},
Steve B. Howell\altaffilmark{3},
Mark Seibert\altaffilmark{4},
Luciana Bianchi\altaffilmark{5}}
\altaffiltext{1}{Department of Astronomy, University of Washington,
Box 351580,
Seattle, WA 98195, szkody@astro.washington.edu,plotkin@astro.washington.edu}
\altaffiltext{2}{Department of Astronomy, New Mexico State University, Box 30001, MSC 4500, Las Cruces, NM 88003, tharriso@nmsu.edu}
\altaffiltext{3}{WIYN Observatory and National Optical Astronomy Observatories,
950 N. Cherry Avenue, Tucson, AZ 85726, howell@noao.edu}
\altaffiltext{4}{California Institute of Technology, MC 405-47, Pasadena, CA 91125, mseibert@srl.caltech.edu}
\altaffiltext{5}{Center for Astrophysical Sciences, The John Hopkins University,
3400 N. Charles St., Baltimore, MD 21218, bianchi@skysrv.pha.jhu.edu}

\begin{abstract}
Low state optical photometry of EF Eri during an extended low accretion state
combined with GALEX near and far UV 
time-resolved photometry reveals 
a source of UV flux that is much larger than the
underlying 9500K white dwarf, and that is highly modulated on the orbital period.
The near UV and
optical light curves can be modeled with a 20,000K spot but no spot model can 
explain both
the large amplitude FUV variations and the SED. The limitations of limb
darkening, cyclotron and magnetic white dwarf models in explaining the 
observations are
discussed.
\end{abstract}


\section{Introduction}

Accreting close binaries containing white dwarfs with high (10-200 MG)
magnetic fields are termed AM Herculis variables or simply polars (review
in Wickramasinghe \& Ferrario 2000). They tend to have lower mass transfer
rates than disk accreting systems of similar orbital period and spend perhaps
the majority of their time (Ramsay et al. 2004)
 in states where the mass transfer can drop by orders of magnitude 
 or stop
completely (termed low states). During these times, most of the effects
associated with active accretion (X-rays, strong optical cyclotron continuum,
strong He and Balmer emission lines) disappear. This provides the opportunity to
directly view and study the underlying stars (white dwarf and late type secondary).
The polar cataclysmic variable EF Eri received notoriety
as it remained in one of these inactive states for 9 years, starting in 1997
(Wheatley \& Ramsay 1998; WR98) and only becoming active again at the beginning of
2006.  In the high state, EF Eri is 14th mag and shows a highly variable
light curve over the 81 min orbital period (Bailey et al. 1982). In the
infrared, the J light curve revealed a narrow dip which Bailey et al. used
to define  phase 0.0. At this phase, the $UBV$ light curves showed a broad 
minimum, 
the X-ray had a sharp minimum and circular polarization was
a maximum. These features were interpreted as an eclipse of the accretion
column by the accretion stream at phase 0. Cyclotron humps led to estimates
of 16.5 and 21 MG for two cyclotron regions.

When EF Eri entered its prolonged low state in 1997, the V mag dropped to
18, broad Balmer absorption lines from the white dwarf were apparent, and
Zeeman splitting indicated a magnetic field of 14 MG (WR98). While 
Euchner et al. (2003) fit multipole models (5 components with fields up to
100MG) to spectra obtained in 2000, the fit was not definitive. WR98 concluded
the 
similarity of the low state field
to that measured in the small accretion regions during the high state
indicated either a uniform field strength over the white dwarf, or continued
accretion heated spots.
Beuermann et al. 
(2000) modeled the optical spectrum with a
9500K white dwarf with a 15,000K hot spot covering 6\% of the surface.
$BVRI$ photometry obtained by Harrison et al. (2003; H03) in 2001, showed 
sinusoidal light curves with a peak near phase 0.9 and a minimum near
phase 0.4. They modeled the light curves with a 12,000K spot covering 6\% of
the area of a 9500K white dwarf,
with a best fit obtained for an inclination of 35$^\circ$ and an angle of 35$^\circ$ between the spin
and magnetic pole. The $H$ and $K$ light curves were anti-phased from the 
optical, with a peak near
phase 0.5 and a minimum near phase 0. IR spectra obtained in 2002 (Harrison
et al. 2004; H04) showed different cyclotron harmonics were present at phase 0 
than at phase 0.5, indicating that cyclotron emission was still present from both
accretion poles.  

Given that the optical spectrum of EF Eri at the low state shows no
obvious sign of an accretion stream, it
is intriguing that a hot spot model still fits the optical
light curves and cyclotron is still present in the IR. During a study of polars at low states, Araujo-Betancar (2005) found that 30,000-70,000K hot spots
contributing 20-30\% of the white dwarf flux were needed to model the FUV fluxes. To
further explore the heating effects at extremely low accretion rates,
we obtained UV light curves with GALEX. Our results show the value of GALEX
time-resolved photometry in providing information confirming
that some area of the white dwarf contributes a large amount of
UV flux even after 7 years of
extremely low mass transfer.

\section{Observations}

GALEX observations were first obtained on 2004 November 8 during 10
contiguous satellite orbits from 04:32:38 to 19:47:50 UT. 
The {\it GALEX} satellite (Martin et al. 2005) uses a dichroic to split the
UV light into a FUV detector (1350-1750\AA) and a NUV detector (1750-2800\AA).
The field of view is 1.25$^\circ$ with 5 arcsec resolution in the FUV and 6.5 arcsec
in the NUV. During this observation, elevated solar proton levels resulted
in the FUV detector being switched off, so only data in the NUV were obtained.
The observation was rescheduled and the full 10 orbits with both detectors
were accomplished on 2004 December 7 from 03:46:02 to 18:59:40 UT. 

While the standard GALEX data pipeline converts time-tagged photons into
 a final calibrated image for each observation,
the production of light curves required the generation of calibrated images in
120s intervals\footnote{Produced by the GALEX Software Operations and Data 
Analysis team.}, which were then 
phased according to the
ephemeris of Bailey et al. (1982).
The IRAF\footnote{{IRAF (Image
 Reduction and Analysis
Facility) is distributed by the National Optical Astronomy Observatories, which
are operated by AURA,
Inc., under cooperative agreement with the National Science Foundation.}}
routine {\it qphot} was then used to obtain a magnitude for the photons inside 
a 9 pixel aperture
at the source position. The background was
measured in an annulus of width 3 pixels around the aperture. The
conversions from cps to magnitude to flux were accomplished using the
values from the GALEX online documentation\footnote{
http://galexgi.gsfc.nasa.gov/tools/index.html}
where FUV m$_{0}$=18.82=1.40$\times$10$^{-15}$ ergs cm$^{2}$ s${-1}$ \AA$^{-1}$
and NUV m$_{0}$=20.08=2.06$\times$10$^{-16}$ ergs cm$^{2}$ s${-1}$ \AA$^{-1}$. 
To insure that the optical light curve had not changed from 2001 December
(H03), further photometry was obtained on 2005 September 16.
As in 2001, the New Mexico State University 1.0 m telescope was used with
a CCD and Johnson-Cousins V filter. A differential light curve was obtained
with respect to nearby stars on the same frames and Landolt standards were
used to calibrate the reference stars.
The combined GALEX and V light curves are shown in Figure 1.

\section{Light Curve Modeling with a Hot Spot}

As shown in Figure 1, the FUV, NUV, and $V$ light curves show sinusoidal-like 
variations that appear to have similar phasing to each other and to the
$BVRI$ light curves of H03. The symmetrical nature and width of the variation
implies large and symmetrical regions of emission. 
As with the optical light curves of H03, 
they are all anti-phased to the low state 
$H-$ and $K$-band light curves shown in H04 (the $J$-band light curve is more 
complex). The amplitude of the variations in the optical are 0.2 mag peak-to-peak, while 
the UV and IR variations have considerably larger amplitudes (up to 0.8 mag in
FUV and $K$). The time-resolved IR spectra of H04
confirmed that the $H-$ and $K$-band variations were due to cyclotron 
emission. 

H03 showed the optical variations can be easily modeled using a simple hot spot 
on a 
9500 K white dwarf. They  
used the light curve modeling program WD98 (Wilson 1998) to confirm that 
such a model could explain both their $BVRI$ light curves and the optical 
spectral energy distribution (SED). Presumably, this hot spot would be located 
near one of the magnetic poles, and represent the heating induced by continued 
accretion whose presence is clearly demonstrated by the strong cyclotron 
emission observed during this low state.

Given their identical phasing, we attempted to use the same hot spot model
to explain the GALEX light curves of EF Eri. However, it soon became apparent
that
a 14,000 K hot spot on a 9,500 K  white dwarf does not provide sufficient 
ultraviolet flux at short wavelengths to simultaneously explain the observed
NUV and FUV fluxes. To achieve a reasonable fit to the observed FUV/optical
SED requires that the hot spot have a temperature of T$_{\rm eff}  \geq$ 20,000 
K. Figure 2 shows the spectral energy distributions (SEDs) for the GALEX and
optical data for
minima (phase 0.4; filled dots) and maxima (phase 0.9; stars) light (the values used are listed in Table 1). The
fit to the minimum SED (green line) is for a 9800K white dwarf (red dotted line)
that contributes 78\% of the V flux, with the rest from a 22,000K hot spot
(blue dotted line). It is obvious that to account for the light curve 
maxima will require a larger contribution of a component near 20,000K. 

With this insight, we constructed a large number of models using 
WD2005\footnote{WD2005 is an updated version of WD98, and can be obtained at 
this website maintained by J. Kallrath: 
http://josef-kallrath.orlando.co.nz/HOMEPAGE/wd2002.htm} with differing spot 
sizes, orbital inclinations and spot latitudes. For our trial runs, none of our 
models included limb darkening.  
 Our 
initial models had a single spot temperature. Given the large number of 
adjustable parameters, it was easy to generate a model that could fit the
$V$ and NUV light curves.
Figure 3 shows a model
for an inclination of 45$^\circ$ and a 20,000K spot with  
a radius of 60$^\circ$
located 10$^\circ$ from the spin pole. However, no single temperature spot model
could explain the large amplitude of the FUV light curve--even unrealistic
models where one entire hemisphere was at 20,000 K, and those types of models
could not simultaneously explain the UV/optical SED. 

The difficulty with the single temperature spot models is that they must 
supply both an enormous FUV flux $and$ produce $~\sim$ 1 magnitude variations. 
For example, a 9500 K white dwarf normalized to the $V$-band flux, 
only supplies about 4\% of the observed FUV flux. Thus, this 20,000K spot must 
provide the remaining flux in the FUV at all orbital phases, and somehow be 
highly variable. We thus explored more complicated models, ones in which the 
temperature of the ``core'' of this spot was very high (T$_{\rm eff}$ = 100,000 
K), which had cooler, off-centered annuli surrounding it that smoothly dropped 
off in temperature with radius until they reached the observed photospheric 
temperature of the white dwarf. Even these models were unable to produce both
the large amplitude FUV variations and the observed SED.

One set of spot models that could reproduce the observed
light curve variations was models with extreme FUV limb darkening. Using a 20,000 K
spot, models with extreme limb darkening in the FUV (i.e. a coefficient of 2.0, a 
factor of 2 over normal values for hot stars),
 while the NUV
and V-band limb darkening coeffients had normal values of 0.5 and 0.3 
(Al-Naimiy 1987, Wade \& Rucinski 1985) could produce large amplitude variations. 
Figure 4 shows this type of model fit to the GALEX and V data. 
While we cannot fully rule out such 
models, such an abrupt change in limb 
darkening laws between the FUV and NUV bandpasses is unlikely, especially given 
the 
expectation of few large spectral line features in the GALEX bandpasses. It is 
interesting to note that Diaz et al. (1996) found quite large limb darkening
effects for accretion disks in the UV, depending on the line-of-sight 
inclination angle of the disk. A tilted, flat surface like an accretion disk 
has larger limb darkening than a spherically symmetric star since the 
emergent flux from a spherical object is averaged over all angles. Thus, it does
not seem impossible to envision a larger FUV limb darkening for a relatively 
small, isolated spot. If one were to add some vertical extent to this spot, its 
limb darkening could be quite large. But, no matter what limb darkening is used, such a 
spot/region needs to have a spectrum that preferentially produces FUV emission.

\section{Discussion}

While it is now evident that a source of far UV photons remains on the white dwarf during
extremely low states of accretion, the nature of this component is not clear. 
G\"ansicke et al. (2006) were able to successfully fit far UV (FUSE) and near UV (STIS)
light curves of AM Her during a low state using a hot spot of 
34,000-40,000K on a 20,000K white
dwarf. In comparison to EF Eri, their bandpass closest to our
GALEX FUV one shows amplitudes that are about half those for EF Eri and the 
hotter
white dwarf and spot can account for the peak amplitudes and fluxes near
1100\AA. However, they did not see any change in temperature of the hot
spot in AM Her over the course of several months, and puzzled over the possible
sources of heating such as low ongoing accretion during the low state 
or deep heating of the
pole cap area during times of high accretion. 

For EF Eri, the results of H04 and the shapes of the light curves offer some
clues to other possibilities. The IR cyclotron features at the low state indicate two poles with different
magnetic fields (if other poles are present, they are weak compared to these). 
The opposite behavior of J vs H and K indicates J has a
cyclotron feature that is likely from a different accreting pole during the
high state (evident at photometric phase 0), whereas the low state shows cyclotron from the pole evident near phase 0.5. It is interesting that both the GALEX and V
data (Figure 1) show a disturbance in the light curves near phase 0.5, that
could indicate cyclotron is somehow playing a role in the UV. The
J band light curve at the low state also shows a slight peak at this phase.
For field strengths such as present in EF Eri, the cyclotron harmonics are
usually optically thick in the IR and optically thin in the optical and UV.
The thin harmonics would be best seen perpendicular to the magnetic field,
thus explaining the offset in phase from the opt/UV to the IR, if cyclotron
was responsible. 

The optical light curves of the Low Accretion Rate Polars (LARPS) show very similar 
sinusoidal variability (Szkody et al. 2003) that can be explained by the
presence of cyclotron harmonics in the bandpass of observation. Thus, if 
there were a cyclotron component in the UV, it could account for the
extra UV light. However, for the field strengths listed, it would be
difficult to get a component only in the UV that is not present in the
optical. 
Schwope, Schreiber \& Szkody (2006) have proposed cyclotron components in
the UV to explain the discrepancy between the SED and model WD in the polar
RXJ1554.2+2721. That system has a field strength of 110 MG and
shows noticeable cyclotron humps
in the optical. While Euchner et al. (2003) proposed multipoles up to these
field strengths for EF Eri, their spectra showed no visible cyclotron humps.
However, a recent Keck spectrum taken in 2006 January, as EF Eri was emerging
from its long low state (Howell et al. 2006), shows possible harmonics near 
9300\AA\ and 4700\AA\
which are consistent with a 115 MG field.

Schmidt et al. (2005) point out the discrepancy in LARPS between the
higher observed blue flux compared to white dwarf models that fit the longer 
wavelengths.
This same discrpancy was also shown in modeling the UV flux of the highest field
polar AR UMa (G\"ansicke et al. 2001) and ascribed to the lack of known
correct models for high magnetic field white dwarfs. Thus, it is possible
that the discrepancy could go away if realistic models show a steeper
flux distribution for the white dwarf areas near the magnetic poles.
Modeling also suffers from correct treatment of the irradiated atmospheres of
white dwarfs. K\"onig, Beuermann \& G\"ansicke
(2006) have recently explored the atmosphere around a polar cap 
irradiated by bremsstrahlung and cyclotron flux. For low magnetic field systems,they find the irradiated areas will be large and the reprocessed 
energy will be in the FUV. While their models are for higher accretion rates
than in systems like EF Eri and LARPS, and they point out
that a much larger effort will be required to model the lowest accretors
correctly, it is a step in the right direction.

Whether the extra
UV flux is due to hot spots, cylotron harmonics or incorrect UV models
for high field white dwarfs
remains to be determined. 

\section{Conclusions}
Our results show that GALEX time-resolved photometry is a very useful
means of studying the hot components in close binaries.
Our near and far UV light curves obtained during the extended low
accretion state of EF Eri reveal the presence of a symmetrical, large and
highly variable source of
UV light. If this is a
hot spot on the 9500K white dwarf, this spot 
must have a temperature $>$20,000K to fit the SED. However, the amplitude
of the FUV variation is far too large to be fit with a spot of this type,
unless large limb darkening is invoked just for the FUV. Since the UV light
curves are in phase with the optical, and opposite in phase from IR
variations that are known to originate from cyclotron harmonics, it is natural 
to invoke changing optical depth of harmonics from optically thick in the 
IR to optically thin in the optical/UV. If there was a harmonic present in
the far UV band, it could explain the excess FUV flux and amplitude, but this
would require extremely high fields and harmonics should be visible in the 
optical as well. It is intriguing that small deviations
in the UV and optical light curves match the phases of peak IR flux and that the
excess UV luminosity over that of the 9500K white dwarf is comparable to the
cyclotron luminosity in the IR. However, the lack of good models for 
cyclotron effects on the spectrum of a white dwarf 
prevents us from reaching any definite conclusion.
 Until we can correctly model the
observed excess UV flux in the cases of high magnetic field white dwarfs, 
we cannot ascertain its cause as a hot spot or cyclotron emission or
atmospheric effects due to high fields and low accretion.

\acknowledgments

Support for this research was provided by NASA GALEX grant NNG05GG46G.

\clearpage

\clearpage
\begin{deluxetable}{ccc}
\tablewidth{0pt}
\tablecaption{Mean Mags at Bright \& Faint Phases}
\tablehead{
\colhead{Wavelength(\AA)} & \colhead{Phase 0.9} & \colhead{Phase 0.4}}
\startdata
1528 & 18.58 & 19.30 \\
2271 & 18.37 & 18.70 \\
4360 & 18.04 & 18.37 \\
5450 & 18.04 & 18.21 \\
6380 & 17.90 & 18.08 \\
7970 & 17.76 & 17.94 \\
\enddata
\end{deluxetable}

\clearpage
\begin{figure} [h]
\figurenum {1}
\plotone{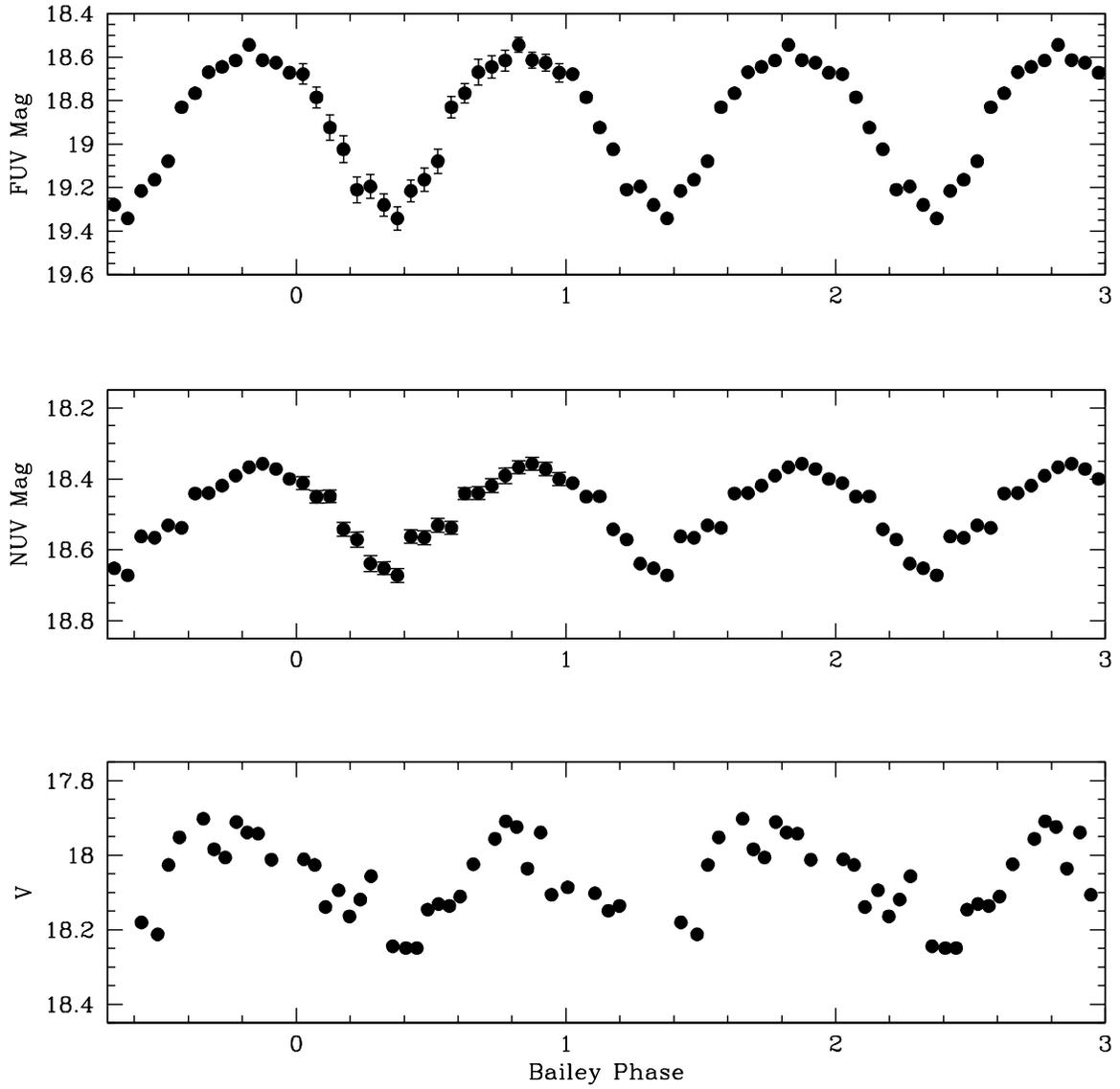}
\caption{GALEX FUV (top), NUV (mid) and V (bottom) light curves as a function of
Bailey phase. Error bars are shown for one cycle in the GALEX plots and
the V light curve shows 2 sequential orbits which are repeated. Error bars
in the V data are $\pm$0.07mag.}
\end{figure}

\begin{figure}
\figurenum {2}
\plotone{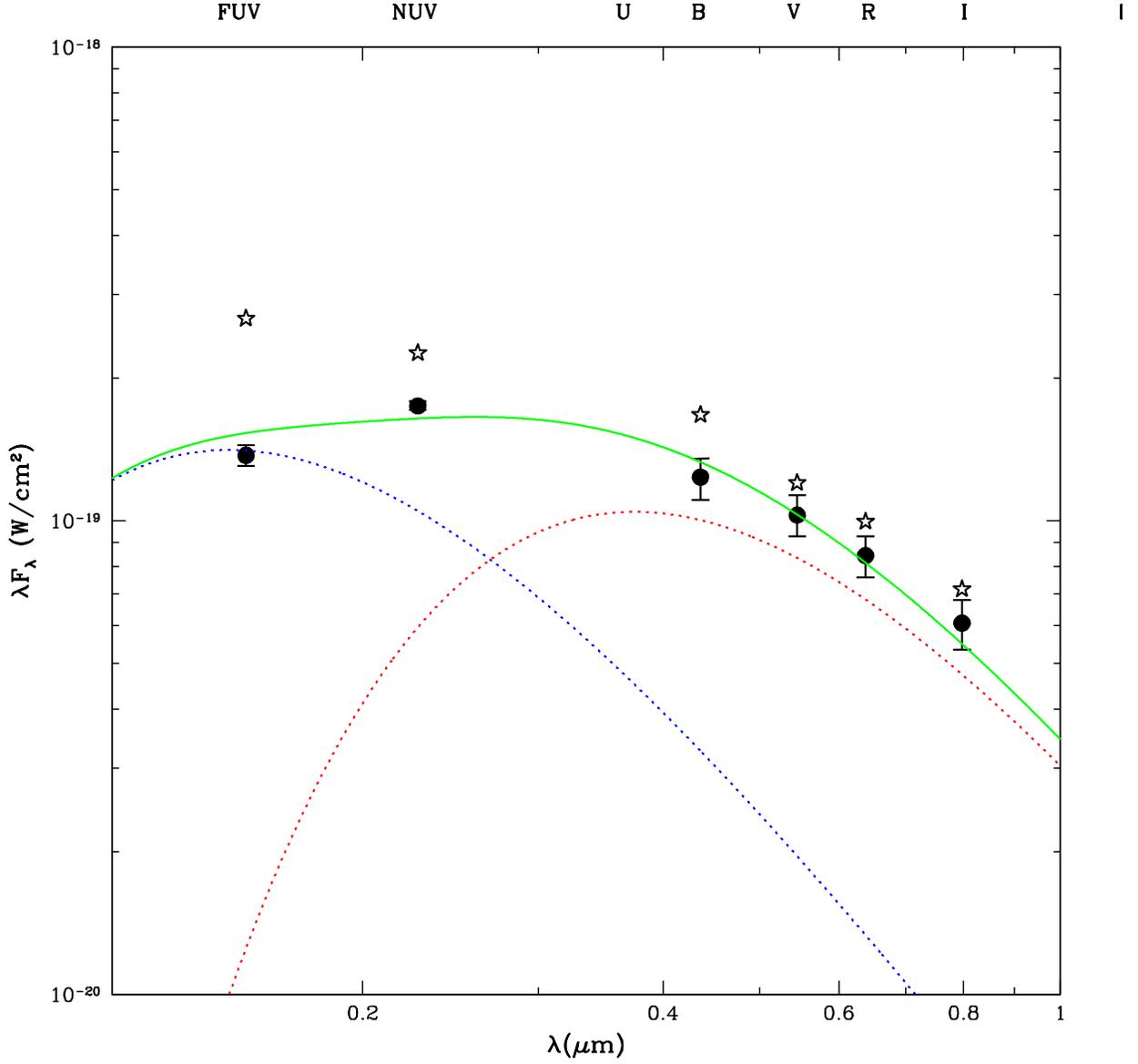}
\caption{SED for GALEX through optical wavelengths. Filled circles are
minima light (phase 0.4), stars are maxima light (phase 0.9), solid green line is
the sum of a fit of a 9800K blackbody contributing 78\% of the V flux (right red
dashed curve) and a 22000K blackbody (left blue dashed curve). Note that the
fit of the maxima light would require a larger contribution of the 22000K BB.}
\end{figure}

\begin{figure}
\figurenum {3}
\plotone{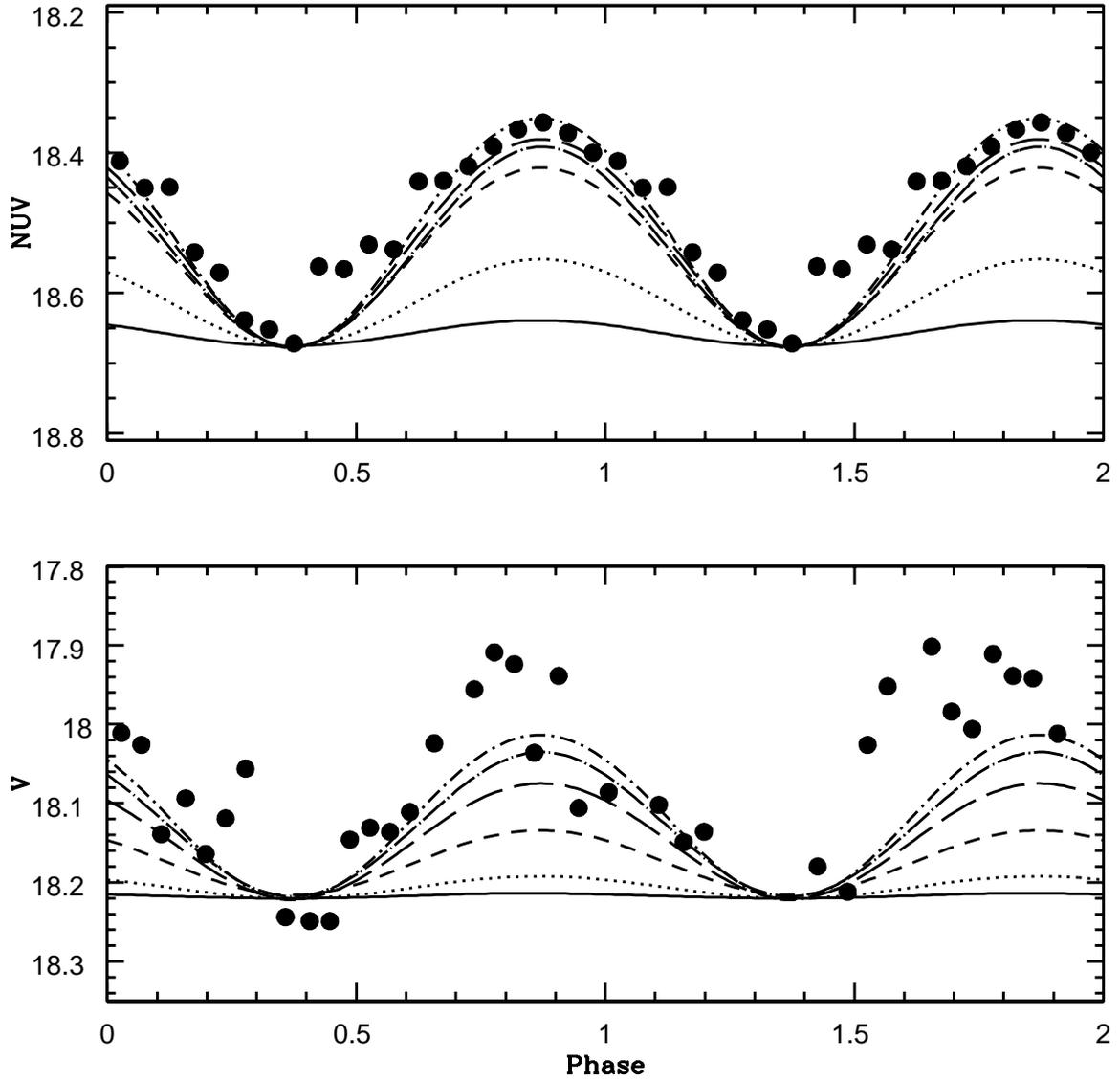}
\caption{Best fit spot model to the GALEX NUV and optical V light curves 
with a 9800K white dwarf and a 20,000K hot spot located 10$^{\circ}$ from
the spin axis with an inclination of 45$^{\circ}$. Solid, dotted, short
dash, long dash, short dash-dot, and long dash-dot curves correspond to
spot radius of
5,10,20,30,45 and 60$^{\circ}$.}
\end{figure}

\begin{figure}
\figurenum {4}
\plotone{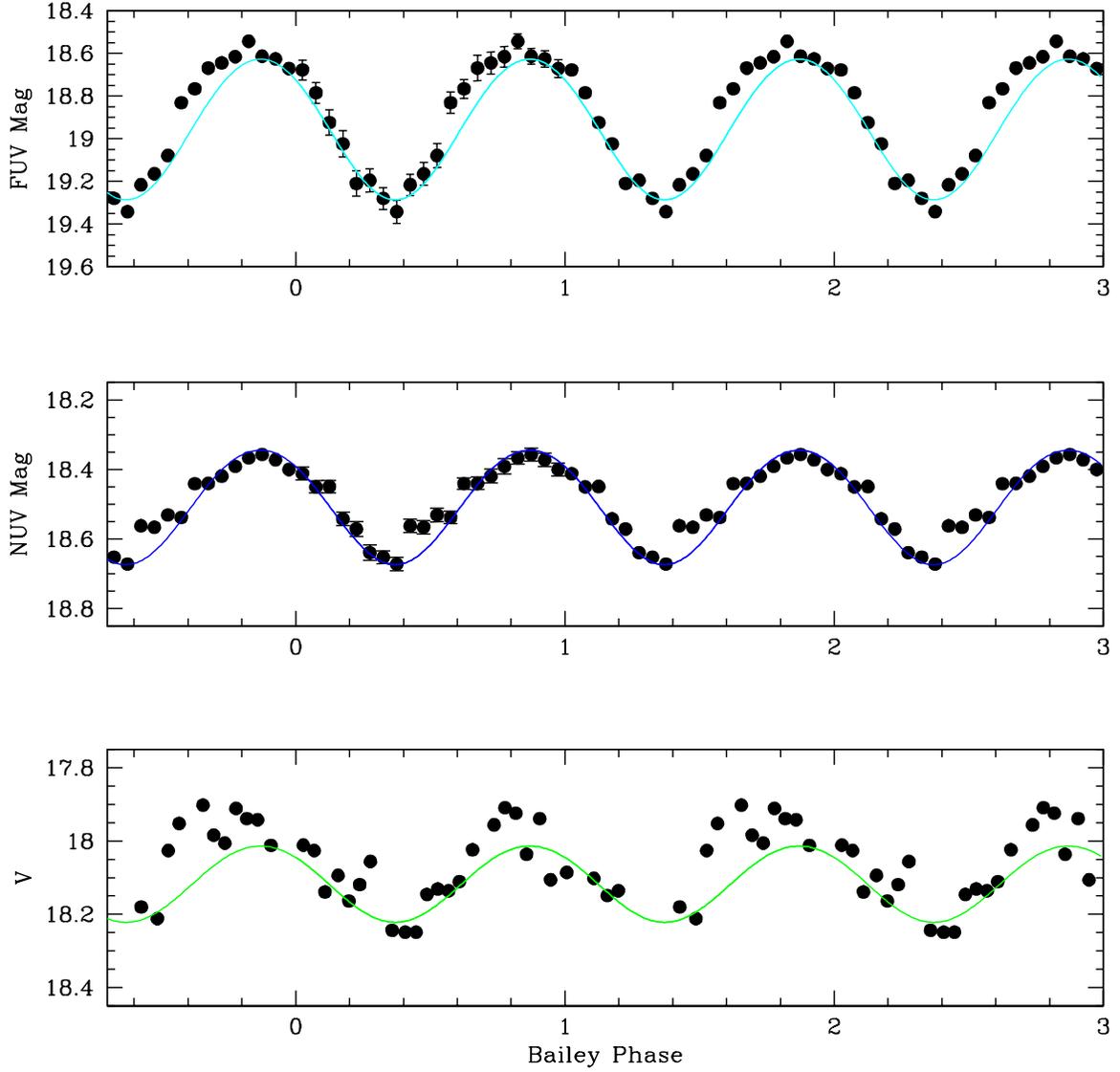}
\caption{Best fits to GALEX and V light curves with a 20,000K spot model and
a large FUV limb darkening (limb darkening coefficients of 2.0, 0.5 and 0.3 for
FUV, NUV and V).}
\end{figure}

\end{document}